**Title**: ARIMA forecasting of COVID-19 incidence in Italy, Russia, and the USA

Gaetano Perone[1]

**Abstract**

The novel Coronavirus disease (COVID-19) is a severe respiratory infection that officially occurred in Wuhan, China, in December 2019. In late February, the disease began to spread quickly across the world, causing serious health, social, and economic emergencies. This paper aims to forecast the short to medium-term incidence of COVID-19 epidemic through the medium of an autoregressive integrated moving average (ARIMA) model, applied to Italy, Russia, and the USA The analysis is carried out on the number of new daily confirmed COVID-19 cases, collected by Worldometer website. The best ARIMA models are Italy (4,2,4), Russia (1,2,1), and the USA (6,2,3). The results show that: i) ARIMA models are reliable enough when new daily cases begin to stabilize; ii) Italy, the USA, and Russia reached the peak of COVID-19 infections in mid-April, mid-May, and late May, respectively; and iii) Russia and the USA will require much more time than Italy to drop COVID-19 cases near zero. This may suggest the importance of the application of quick and effective lockdown measures, which have been relatively stricter in Italy. Therefore, even if the results should be interpreted with caution, ARIMA models seem to be a good tool that can help the health authorities to monitor the diffusion of the outbreak.

**Keywords**: COVID-19; epidemic; ARIMA; forecasting; time series; public health.

**JEL Codes**: I10; I18; C22; C53.

**1. Introduction**

Coronavirus disease (COVID-19) is a severe acute respiratory syndrome that occurred for the first time in Wuhan, the capital city of Hubei Province in China, in December 2019. In late February, the virus was detected even in Europe, and from that moment on, it has continuously spread across the world. As of 31 May, according to Worldometer's COVID-19 data, the pandemic affected 213 countries and territories, and two conveyances, with more than 6,2 million confirmed cases and a death toll higher than 370,000 people. The main epicenters of the pandemic are mostly advanced countries, i.e. Brazil, France, Germany, Italy, Russia, Spain, the UK, and the USA. However, in the last few weeks, other relatively poorer countries, such as India, Peru, and Turkey, are progressively climbing the world rankings.
When an epidemic occurs, one crucial issue is determining its evolution and inflection point. Therefore, the aim of this paper is to provide a short to medium-term forecast of the spread of the COVID-19 disease, and its inflection point, in three of the most affected countries worldwide, Italy, Russia, and the USA. The prediction will be estimated by using an autoregressive integrated moving average (ARIMA) model, applied to the number of new daily confirmed positive cases of COVID-19. The specific time frame for each country will be chosen by considering, like departure point, the moment when daily cases began to show signs of stabilization.

---

[1] University of Bergamo. E-mail: gaetano.perone@unibg.it.

For the remainder, this paper is organized as follows. In section 2, I will introduce ARIMA model and related literature. In section 3, I will present the data used to forecast and discuss the empirical strategy. In section 4, I will discuss the results. In section 5, I will provide some concluding considerations and possible policy implications.

## 2. Arima models and related literature

In the last few months, an increasing body of literature has attempted to forecast the incidence and prevalence of the COVID-2019 pandemic by using different approaches, such as the ARIMA (Benvenuto et al., 2020; Chakraborty and Ghosh, 2020;[2] Ribeiro et al., 2020[3], Singh et al., 2020), the exposed-identified-recovered (EIR) model (Xiong and Yan, 2020), the susceptible-exposed-infected-recovered (SEIR) model (Wu et al., 2020), the segmented Poisson model (Zhang et al., 2020), the SIDARTHE[4] model (Giordano et al., 2020), the susceptible-infected-recovered (SIR) model (Nesteruk, 2020), the SIR/death model (Anastassopoulou et al., 2020; Fanelli and Piazza, 2020), and mathematical methods based on travel volume (Tuite et al., 2020).

The ARIMA model is clearly one of the most preferred because of its good properties. In fact, it is quite easy to fit, manage, and its mathematical interpretation is easy and immediate also for non-academics.

It was introduced for the first time, as the Box-Jenkins approach, by statisticians Box and Jenkins in a highly influential seminal work published in 1970. It could be considered one of the most used prediction models for epidemic time series (Rios et al., 2000; Li et al., 2012; Zhang et al., 2014; Zheng et al., 2015), and it is frequently used with non-stationary time series to capture the linear trend of an epidemic or disease. It allows the prediction of a given time series by considering its own lags, i.e. the previous values of the time series, and the lagged forecast errors.

Moreover, it is very flexible and can be easily adapted to any kind of data, considering trend, cyclicity, seasonality, calendar variation, randomness disturbances like other diseases, external or exogenous interventions, outliers, and other relevant real aspects of time series (Pack, 1990; Barnett and Dobson, 2010).

Therefore, the ARIMA models allow in a simple way to investigate COVID-2019 trends, which are currently of huge economic and social impact, by helping the health authority to continuously monitor the epidemic and to better allocate the available resources.

In Table 1, I selected 19 studies on the prediction of the spread of various diseases by using an ARIMA framework.

---

[2] Specifically, the authors developed a hybrid ARIMA-Wavelet-based forecasting (WBF) model.
[3] The authors also used cubist regression (CUBIST), random forest (RF), ridge regression (RIDGE), support vector regression (SVR), and stacking-ensemble learning.
[4] This model considered eight stages of infection: susceptible-infected-diagnosed-ailing-recognized-threatened-healed-extinct.

Table 1. Nineteen selected studies on disease forecasting, that uses an ARIMA approach.

| Authors | Disease | Methodological approach | Investigated area |
|---|---|---|---|
| Earnest et al. (2005) | SARS | ARIMA | Singapore |
| Gaudart et al. (2009) | Malaria | ARIMA, SIRS | Mali |
| Liu et al. (2011) | HFRS | ARIMA | China |
| Li et al. (2012) | HFRS | SARIMA | China |
| Ren et al. (2013) | Hepatitis E | ARIMA, BPNN | Shanghai, China |
| Kane et al. (2014) | H5N1 | ARIMA, Random Forest time series | Egypt |
| Zheng et al. (2015) | Tuberculosis | SARIMA | Xinjiang, China |
| Wei et al. (2016) | Hepatitis | SARIMA, GRNN | Heng County, China |
| Zeng et al. (2016) | Pertussis | SARIMA, ETS | China |
| Xu et al. (2017) | Mumps | SARIMA | Zibo, China |
| He and Tao (2018) | Influenza | ARIMA | Wuhan, China |
| Wang et al. (2018) | Hepatitis B | SARIMA, GM (1,1) | China |
| Cong et al. (2019) | Influenza | SARIMA | Mainland China |
| Wang et al. (2019) | Brucellosis | ARIMA | Jinzhou, China |
| Benvenuto et al. (2020) | COVID-19 | SARIMA | China |
| Ceylan (2020) | COVID-19 | ARIMA | France, Italy, Spain |
| Cao et al. (2020) | Human Brucellosis | SARIMA | Hebei, China |
| Polwiang (2020) | Dengue fever | ANN, ARIMA, MPR | Bangkok, Thailand |
| Singh et al. (2020) | COVID-19 | ARIMA | 15 countries |

Notes: ARIMA, autoregressive integrated moving average; ANN, artificial neuron network, BPNN, back propagation neural network; ETS, exponential smoothing model; GM (1,1), gray model; GRNN, generalized regression neural network, HFRS, hemorrhagic fever with renal syndrome. GM (1, 1), H5N1, highly pathogenic avian influenza; MPR, multivariate Poisson regression; SARIMA, seasonal autoregressive integrated moving average; SIRS, susceptible-infectious-recovered-susceptible.

## 3. Materials and method

The analysis is carried on the number of new daily confirmed positive cases of COVID-19 in Italy, Russia, and the USA, collected by the data aggregator website named Worldometer.[5] As suggested by several authors (Box and Tiao, 1975; McCleary et al., 1980; Box et al., 1994), a reasonable ARIMA model requires at least 40–50 observations. Therefore, to meet this criterion, I considered a minimum of 53 and a maximum of 69 data points.

The specific investigated time window for each country was chosen by identifying the period when daily cases began to stabilize, which can be approximately seen as the peak of the epidemic (Figures 1, 2, and 3). Based on this, I chose the following timeframes: Italy (February 22–April 14), USA (March 9–May 16), and Russia (March 22–May 22). In particular, to forecast the COVID-19 outbreak spread in Italy, Russia, and the USA, I used a nonseasonal ARIMA model. In fact, the autocorrelation function (ACF) and the partial autocorrelation function (PACF) of the raw time series showed a lack of any seasonal significative pattern.[6]

The nonseasonal ARIMA models are generally classified as "AR-I-MA (p, d, q)", where: i) p is the order of autoregressive terms (AR); ii) d is the order of nonseasonal differences (I) required to make the time series stationary; and iii) q is the order of moving average (MA). The optimal parameters for ARIMA models were chosen by considering the following four sequential steps:[7]

i. First, I applied the auto.arima() function, included in the "forecast package" of the widely known R Project for Statistical Computing, developed by Hyndman and Khandakar (2008). This algorithm allows to identify the best order of an ARIMA process by considering a unit root test to identify the appropriate degree of differencing,[8] and the minimization of the corrected Akaike's Information Criterion (AICc)[9] and the log-likelihood function (MLE) to identify the AR and MA parameters;

ii. Second, I evaluated the forecast accuracy by implementing the following four accuracy measures: the mean absolute error (MAE), mean absolute percentage error (MAPE), mean absolute scaled error (MASE), and root mean squared error

---

[5] The data are available at the following URL: https://www.worldometers.info/coronavirus/.

[6] This hypothesis is also confirmed by the auto.arima() function, which will be introduced shortly.

[7] The procedure combines Hyndman and Athanasopoulos (2018) suggestions, as reported in Sections 3.3, 3.4, and 8.7.

[8] In this specific analysis, I applied the KPSS test developed by Kwiatkowski et al. (1992), which is the default method implemented in the auto.arima() function. In fact, the application of the alternative tests, the augmented Dickey-Fuller (1979) (ADF) test and Phillips-Perron (1988) (PP) test, does not change the meaning of the final outcome.

[9] The AICc is a bias-corrected version of the original Akaike information criterion (AIC), proposed by Sugiura (1978) and Hurvich and Tsai (1989), which performs significantly better than the latter in both small and moderate sample sizes, as in this case (Hurvich and Tsai, 1989).

(RMSE).[10] These accuracy measures allow the validation of the models selected through the auto.arima() algorithm.

iii. Third, I examined the autocorrelation function (ACF) and the partial autocorrelation function (PACF) of the residuals from ARIMA models, to detect the presence of significant spikes.

iv. Finally, I verified the fundamental statistical assumptions, homoscedasticity, and serial correlation.[11] The first was controlled by using Engle's Lagrange Multiplier (1982) test for autoregressive conditional heteroskedasticity (ARCH), and the former was detected by implementing the Ljung-Box's (1978) test for serial correlation.

The basic ARIMA forecasting estimated equation is the following [1]:

$$\Delta \hat{y}_t = \lambda_1 \Delta y_{t-1} + .. + \lambda_p \Delta y_{t-p} + \psi_1 \Delta e_{t-1} + .. + \psi_q \Delta e_{t-q} + e_t \qquad [1]$$

Where $\Delta$ means the degree of nonseasonal differences, $\hat{y}$ is the time series that will be predicted at time $t$, $p$ is the lag order of AR, $\lambda$ is the coefficient for each parameter $p$, $q$ is the lag order of MA, $\psi$ is the coefficient of each parameter $q$, and $e_t$ means the residuals of errors in time $t$.

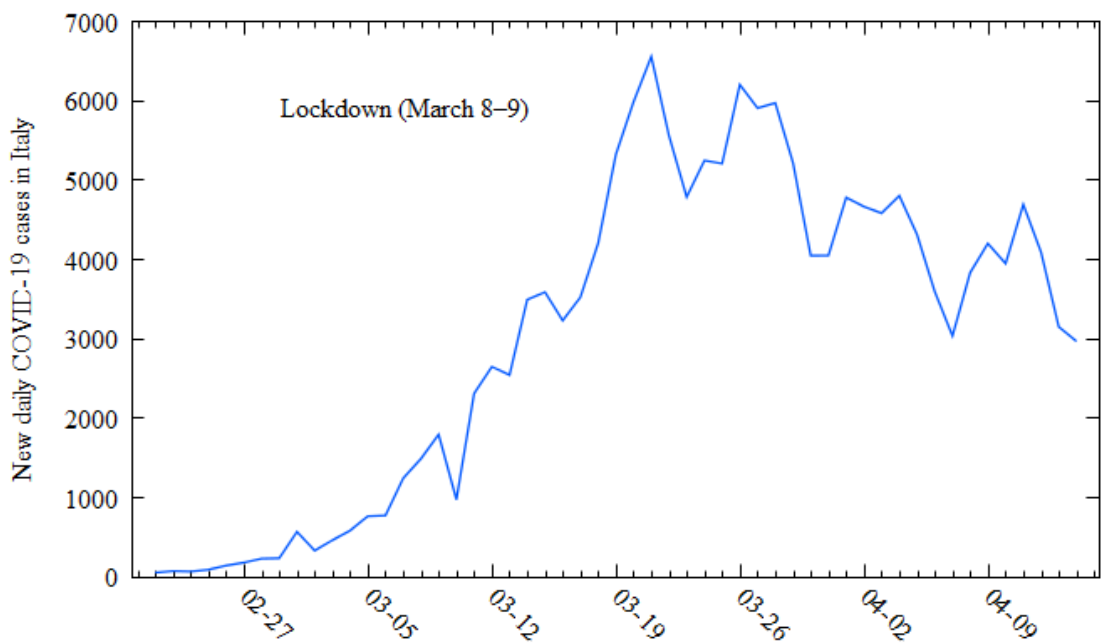

Figure 1. New daily confirmed COVID-19 cases in Italy over the period February 22–April 14, 2020.

---

[10] The mathematical formulas for each measure of forecast accuracy are provided in Table A1 in Appendix A.

[11] The normality assumption of the residuals is not generally necessary (Hyndman and Athanasopoulos, 2018, 3.3).

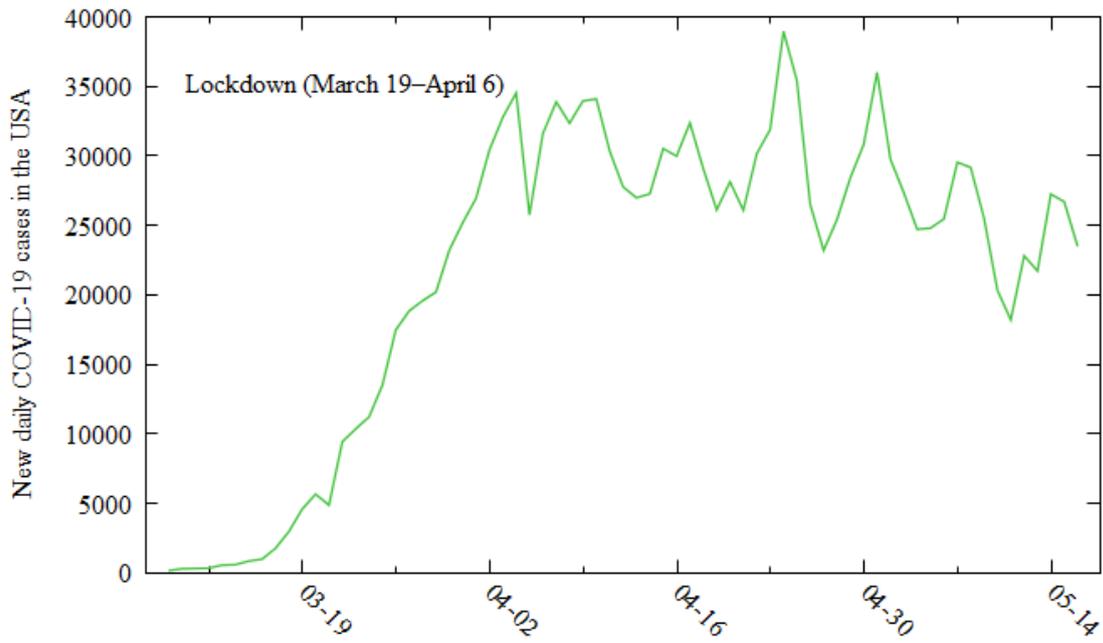

Figure 2. New daily confirmed COVID-19 cases in the USA over the period March 9–April May 16, 2020.

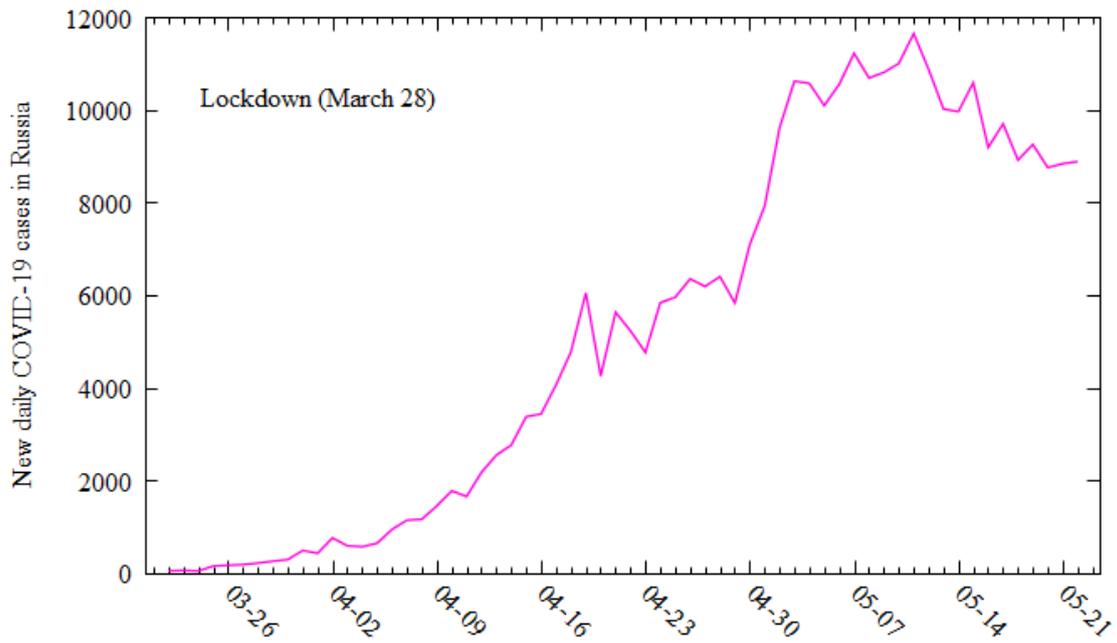

Figure 3. New daily confirmed COVID-19 cases in Russia over the period March 22–May 22, 2020.

## 4. Results and discussion

The minimization of AICc and forecast error measurements suggests the following parameters (Table 2): Italy (4, 2, 4), Russia (1, 2, 1), and USA (6, 2, 3).[12] In the case of Italy, ARIMA (4, 2, 4) is the absolute best according to all criteria. For Russia, MAE, MASE, and RMSE suggest ARIMA (1, 2, 2); however, the AICc and MAPE are definitely lower for the ARIMA (1, 2, 1). For the USA, ARIMA (6, 2, 3) performs far better in all measures of forecast accuracy, but the AICc, which is slightly higher than the ARIMA (2, 2, 4). I chose the former, also based on the residuals auto-correlation structure.

Moreover, MAPE and MASE may give other useful information on the fitting accuracy. In particular, MAPE – which is the most widely used error measure (Goodwin and Lawton, 1999; Ren and Glasure, 2009; Moreno et al., 2013; Kim H. and Kim S., 2016) – indicates an overall forecast accuracy (100-MAPE) of 86.96% for Italy, 88.61% for Russia, and 90.41% for USA.[13] According to Lewis (1982, p. 40), since the values of MAPE are close to 10, the forecasting is definitely good, especially in the case of the USA that shows a value lower than this "limit" (9.59).[14] MASE, which was more recently proposed by Hyndman and Koehler (2008), seems to be more versatile and reliable. Since it is always lower than 1 for all three countries, it indicates that the computed forecast performs definitely better than in-sample one-step forecasts from the naïve method.

The ACF and PACF correlogram plots of the residuals for Italy (Figure C1) and the USA (Figure C3) show a white noise process; in fact, any of the associated spikes go beyond the 95% confidence intervals. Meanwhile, the ACF and PACF correlogram plots for Russia exhibit a significant spike at lag 10, which reaches the 95% confidence limits (Figure C2). Even if it should not be a particular matter of concern, I further examine the ARIMA model process by applying the Ljung-Box's test for autocorrelation and Engle's LM test for the Arch effect.

The Ljung-Box's test shows that the null hypothesis of serial independence of the residuals can be accepted at each distinct lag for all three countries (Table B2). In the same way, Engle's LM test for the ARCH effect shows that the null hypothesis of no autoregressive conditional heteroscedasticity can always be accepted (Table B3). Therefore, the three ARIMA models show a very good fit.

In Figures 4, 5, and 6, I present the forecast results for 30 days in each country. From April 15 to May 14, Italy's coronavirus infection – that is the "benchmark" of the analysis – showed a clear sinusoidal and declining trend (Figure 4). The new daily COVID-19 cases should drop near zero in mid-June 2020, with an estimated epidemic final size of about 246,627 people.

---

[12] The estimated parameters are reported in Table B1 in Appendix B.

[13] It is necessary to stress that the MAPE has also been affected by some criticalities. In fact, it put greater penalty on negative errors than on positive errors and may be very problematic with time points close to zero (Hyndman and Athanasopoulos, 2018, 3.4). However, if the first is mostly out of control, the second aspect is not worry about because all time points considered in the analysis are far from zero.

[14] Specifically, according to the interpretation of Lewis, the models for Italy and Russia are good forecasting, and that for the USA can be considered a highly accurate forecasting.

The outcomes for Russia (Figure 5) and the USA (Figure 6) are also characterized by a declining trend, indicating that they have probably already reached the peak of the COVID-19 outbreak. The major difference between the Italian and Russia/USA cases is that the slowdown in new COVID-19 cases seems to require more time for the latter. Specifically, Russia will reach zero local COVID-19 cases in late August 2020, while in the USA, the new COVID-19 cases will drop to zero between the end of September 2020 and the start of October 2020.

This may confirm the importance of the rapid implementation of strict lockdown measures to contain the spread of the epidemic. In fact, Italy has been under one of the world's strictest lockdown for about 2 months, while Russia and USA have applied easier lockdown restrictions

To control the reliability of my approach, in Tables 3 and 4, I compared the predicted and latest actual daily cases in Italy and the USA, respectively. The results show that in Italy, the overall predicted new daily cases are overestimated, and this deviation grows over time. In the first 10 days of the forecast, the percentage deviation was about 7.27, and it increased until the 30$^{th}$ day of forecast, when it stabilized around 22%. On May 26, i.e. after 42 days, the overall deviation between the predicted and actual values was equal to 15,326 cases, with a change of 22.52 percentage point.

After 10 days, the MAPE was equal to 13.75% and it reached 33.32% at the end of the forecast (after 42 days). Therefore, for Italy, ARIMA performs better at predicting the final size rather than the daily cases.

Regarding the USA, Table 4 shows that the overall predicted cases are a little bit underestimated. In particular, after 5 days, the percentage deviation was about -5.86% and it decreased distinctively after 19 days, reaching 4.11%. After 19 days, the overall deviation between predicted and actual values was equal to 17,128 cases. MAPE, which was equal to 5.73% after 5 days, slightly increased to 5.91% on June 4. Therefore, ARIMA models seem to perform better for the USA than Italy.

About Russia, Table 4 shows that, after 5 days, the percentage deviation was about -2.82%, and it increased to -5.82% after 11 days. MAPE was equal 3.51% after 5 days, and 6.06% after 11 days.

These values can be considered reasonable errors, all things considered.[15] In fact, if we look at Figures 7 and 8, we see that in Italy and the USA, the predicted and actual daily cases follow a very similar and comparable trend. About Russia, the trends seem to differ a little after 5 days (Figure 9). This could indicate that the epidemic is still in the stabilisation phase.

Even if the analytical approach seems to work well enough, these results are a rough guide only and should be additionally verified and updated.

---

[15] In fact, since ARIMA is based on historical observations, the error value in the future is equal to the accumulated random error, one-by-one. Therefore, confidence intervals and errors typically increase as the forecast horizon increases (Hyndman and Athanasopoulos, 2018, 3.5 and 8.8).

Table 2. Comparison of the "best" ARIMA models.

| Countries | AR-I-MA parameters | AICc | MAE | MAPE | MASE | RMSE | $R^2$ |
|---|---|---|---|---|---|---|---|
| Italy | (4, 2, 4) | 787.78 | 283.49 | 13.039 | 0.648 | 412.79 | 0.95 |
| | (4, 2, 2) | 789.28 | 342.96 | 15.341 | 0.784 | 451.28 | 0.95 |
| | (5, 2, 2) | 789.51 | 329.07 | 14.434 | 0.7522 | 439.12 | 0.95 |
| | (4, 2, 5) | 790.1 | 349.69 | 15.801 | 0.6497 | 473.12 | 0.94 |
| Russia | (1, 2, 1) | 947.45 | 430.83 | 11.39 | 0.9338 | 606.66 | 0.98 |
| | (0, 2, 2) | 947.92 | 435.56 | 11.35 | 0.9441 | 609.27 | 0.98 |
| | (0, 2, 1) | 948.73 | 435.68 | 11.81 | 0.9443 | 621.58 | 0.98 |
| | (1, 2, 2) | 949.65 | 426.22 | 11.45 | 0.9238 | 606.01 | 0.98 |
| USA | (2, 2, 4) | 1,262.1 | 2,236.3 | 11.741 | 0.6804 | 3,028.5 | 0.92 |
| | (6, 2, 3) | 1,263.07 | 1,631.3 | 9.59 | 0.6573 | 2,411.6 | 0.95 |
| | (2, 2, 3) | 1,263.73 | 2,236.6 | 11.742 | 0.7554 | 3,028.3 | 0.92 |
| | (6, 2, 1) | 1,263.75 | 1,782.3 | 10.3 | 0.7182 | 2,541.9 | 0.94 |

Notes: parameters (p, d, q); AICc, corrected Akaike's information criterion; MAE, mean absolute error; MAPE, mean absolute percentage error; MASE, mean absolute scaled error; RMSE, root mean squared error; $R^2$, adjusted r-square. The chosen ARIMA models are blue colored.

Table 3. Comparison between the total predicted and actual values in Italy.

| Italy | Values | Values | Values | Values |
|---|---|---|---|---|
| Time window | Until April 24 | Until May 4 | Until 14 May | Until 26 May |
| Overall deviation | +2,217 | +7,953 | + 21,856 | +15,326 |
| Overall % deviation | +7.27% | +16.08% | +21.86% | +22,52% |
| MAPE | 13.57% | 23.17% | 33.4% | 33.14% |
| MAE | 401.01 | 486.76 | 506.27 | 418.47 |
| N. of days | 10 | 20 | 30 | 42 |

Notes: The first day of forecasting is April 15, 2020. The overall deviation was calculated as the total predicted values minus the total actual values (values are rounded). The overall percentage deviation was calculated using the following formula: $[\left(\frac{predicted}{actual} - 1\right) * 100]$.

Table 4. Comparison between the total predicted and actual values in USA.

| USA | Values | Values | Values | Values |
|---|---|---|---|---|
| Time window | Until May 21 | Until May 26 | Until May 31 | Until June 4 |
| Overall deviation | -6,638 | -6,739 | -9,419 | -17,128 |
| Overall % deviation | -5.86% | -3.1% | -2.86% | -4.11% |
| MAPE | 5.73% | 4.46% | 5.07% | 5.91% |
| MAE | 1,390.47 | 1,029.68 | 1,155.37 | 1,332.62 |

| N. of days | 5 | 10 | 15 | 19 |

Notes: The first day of forecasting is May 17, 2020. The overall deviation was calculated as the total predicted values minus the total actual values (values are rounded). The overall percentage deviation was calculated using the following formula: $[(\frac{predicted}{actual} - 1) * 100]$.

Table 5. Comparison between the total predicted and actual values in Russia.

| Russia | Values | Values |
|---|---|---|
| Time window | Until May 27 | Until June 2 |
| Overall deviation | -1,249 | -5,666.18 |
| Overall % deviation | -2.82% | -5.82% |
| MAPE | 3.51% | 6.06% |
| MAE | 318.47 | 546.28 |
| N. of days | 5 | 11 |

Notes: The first day of forecasting is May 23, 2020. The overall deviation was calculated as the total predicted values minus the total actual values (values are rounded). The overall percentage deviation was calculated using the following formula: $\left[\left(\frac{predicted}{actual} - 1\right) * 100\right]$.

Figure 4. New daily COVID-19 cases in Italy (prediction of 30 days).

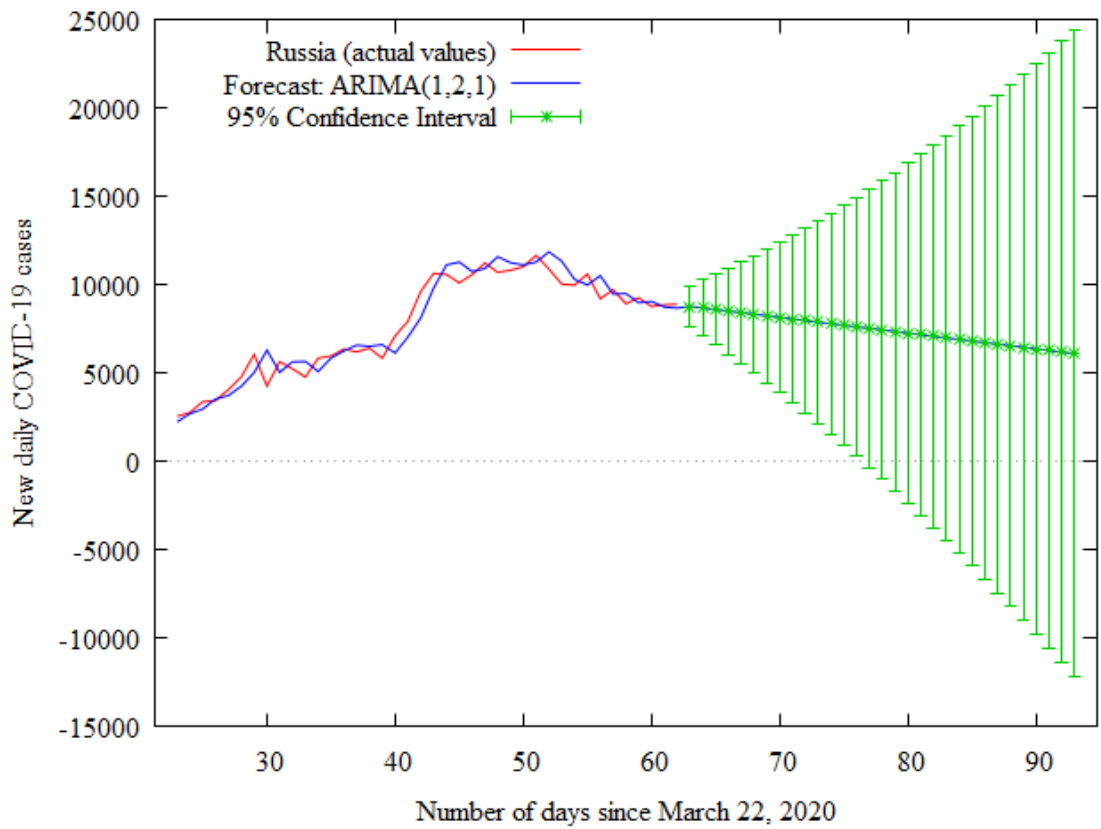

Figure 5. New daily COVID-19 cases in Russia (prediction of 30 days).

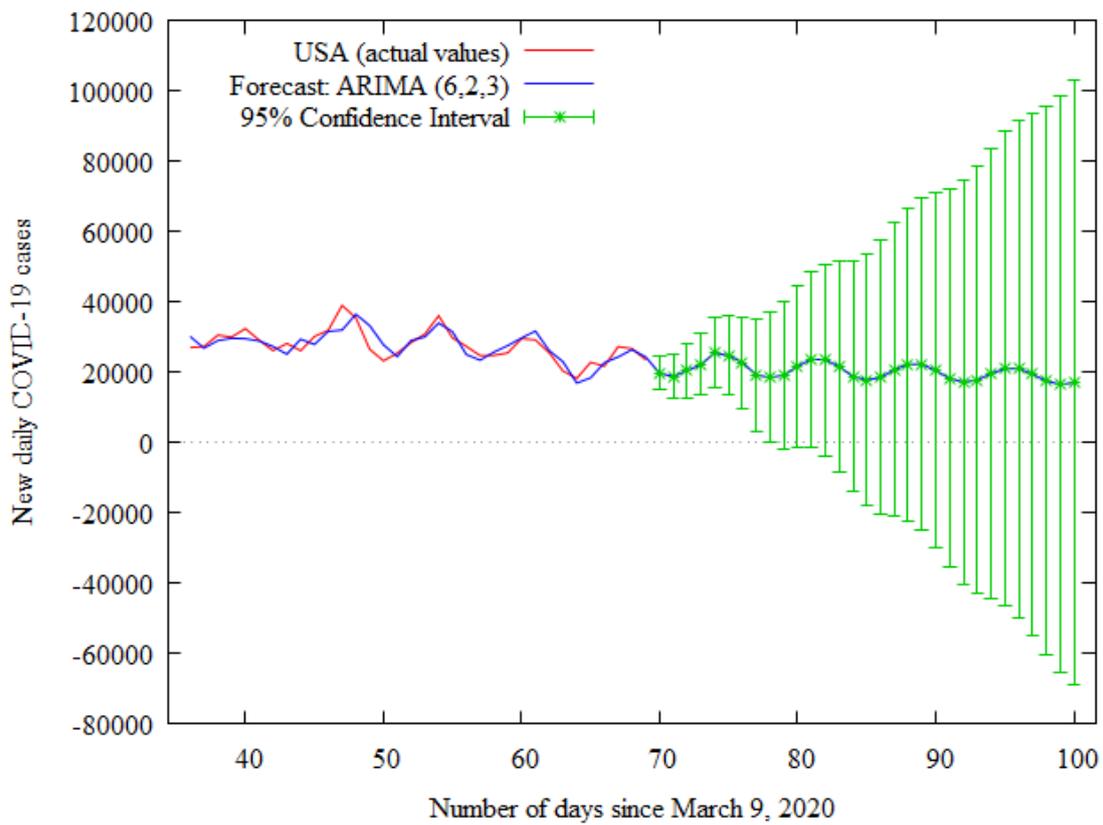

Figure 6. New daily COVID-19 cases in the USA (prediction of 30 days).

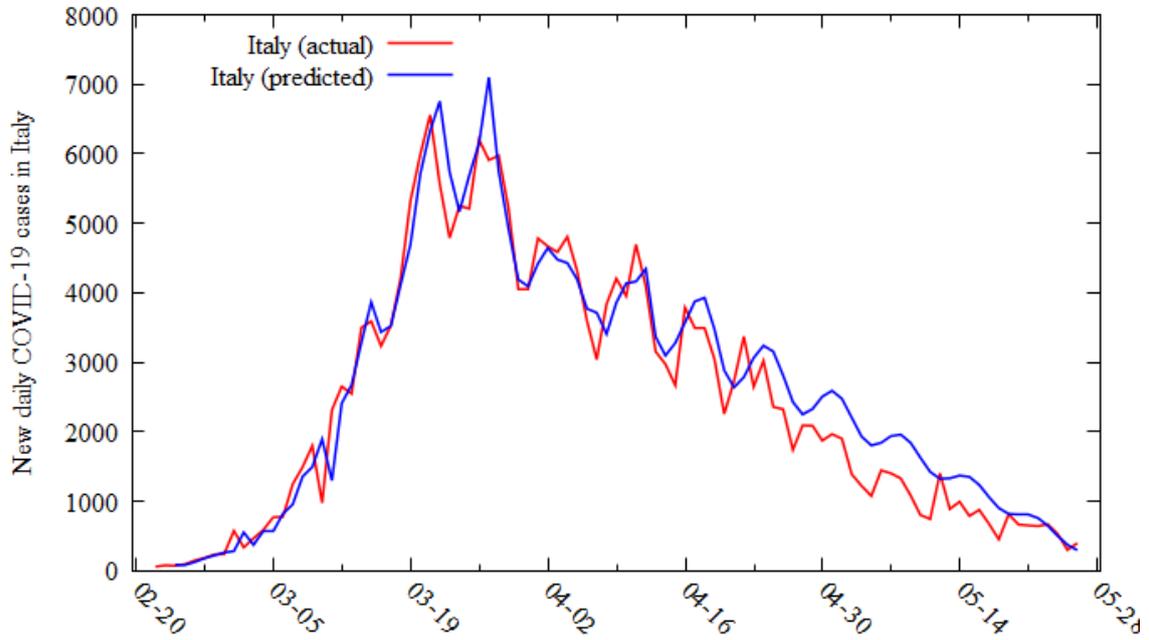

Figure 7. Comparison between actual and predicted values in Italy over the period February 22–May 26.

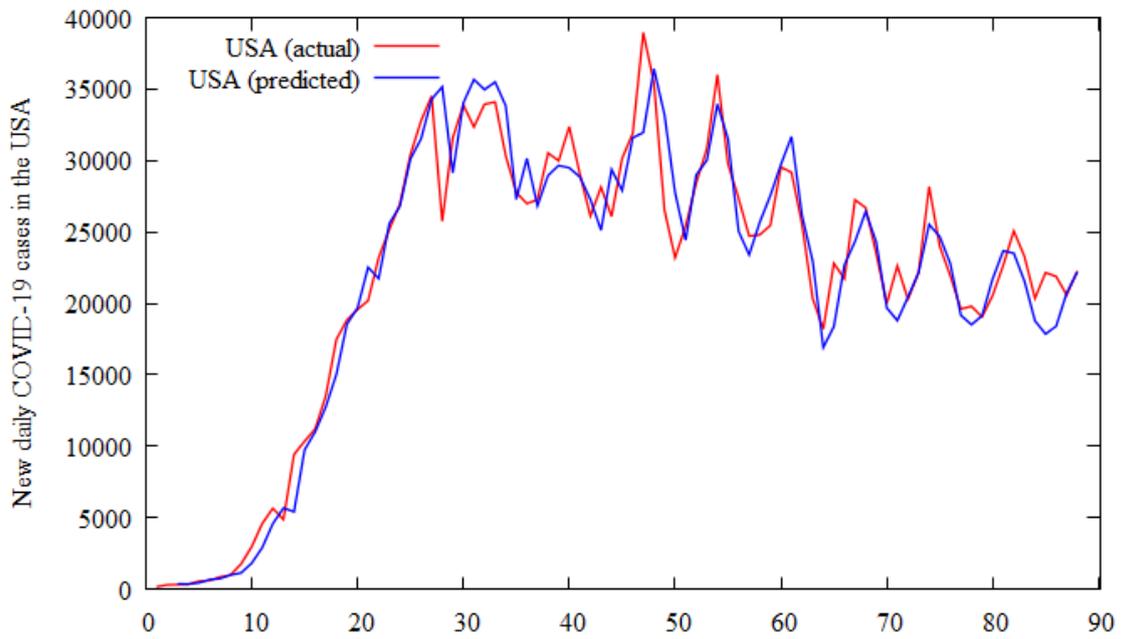

Figure 8. Comparison between actual and predicted values in the USA over the period March 9–June 4.

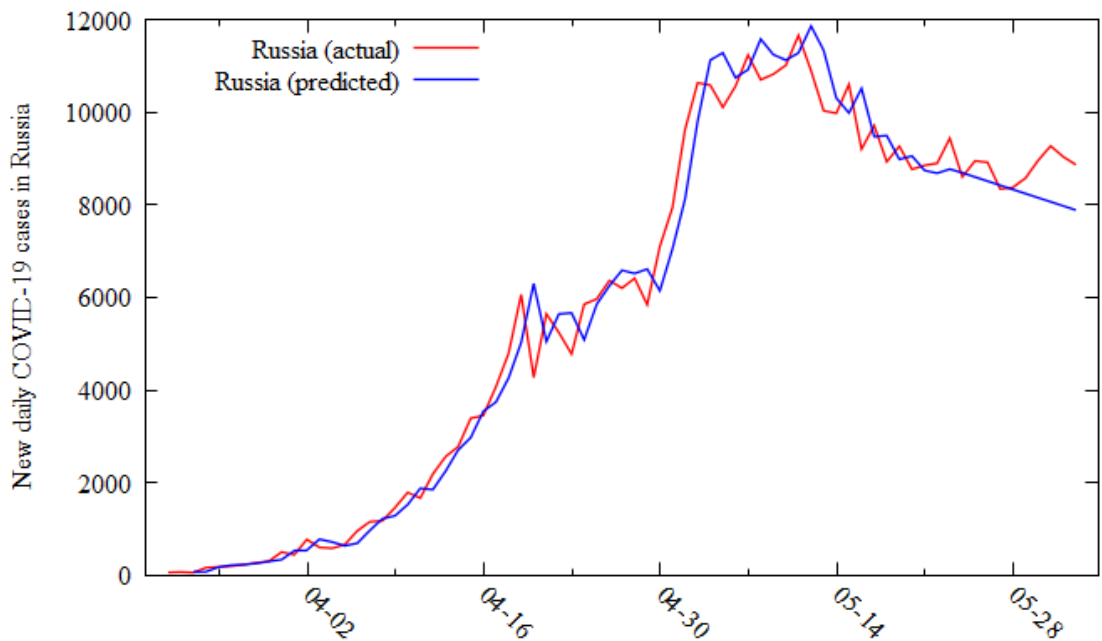

Figure 9. Comparison between actual and predicted values in Russia over the period March 22–June 2.

**Conclusion**

To the best of my knowledge, this is the first study to implement an ARIMA model to predict the incidence of COVID-19 in the short/mid-term in Italy, Russia, and the USA. ARIMA models can be viewed as an easy and immediate tool to program health monitoring systems and to better allocate the available resources. In particular, it can help public health decision makers to plan the number of beds needed for moderate and critical care, and to better allocate and manage medical staff, healthcare devices, and healthcare facilities.

Results suggest that Italy, the USA, and Russia reached the peak of COVID-19 infections in mid-April, mid-May, and late May, respectively. In particular, the USA and Russia will require a considerable length of time to drop near zero daily cases, if compared to Italy. This could be because Italy imposed stricter nationwide lockdown measures such as severe traffic and travel restrictions, bans on gatherings, and closure of commercial activities – to mitigate the spread of the outbreak. Moreover, the comparison between the predicted and more recent actual values showed that forecasts are reliable enough, even in the mid-term, especially for Italy and the USA.

Therefore, ARIMA may be considered a good model for short to medium-term forecasting, but the results should be interpreted with thriftiness. Finally, further useful and more precise forecasting may be provided by continuously updating these data, adding interventions and other real aspects, and applying the model to other countries and/or regions.

# References

Anastassopoulou, C., Russo, L., Tsakris, A., Siettos, C., 2020. Data-based analysis, modelling and forecasting of the COVID-19 outbreak. PloS one, 15 (3), e0230405. https://doi.org/10.1371/journal.pone.0230405.

Barnett, A.G., Dobson, A.J., 2010. Analysing seasonal health data. Springer, Berlin.

Benvenuto, D., Giovanetti, M., Vassallo, L., Angeletti, S., Picozzi S., 2020. Application of the ARIMA model on the COVID-2019 epidemic dataset, Data in Brief, 29, 105340. https://doi.org/10.1016/j.dib.2020.105340.

Bowerman, B.L., O'Connell, R.T., 1979. Time Series and Forecasting: An applied Approach. Duxbury Press, Boston, Massachusetts.

Box, G., Jenkins, G.M., 1970. Time Series Analysis Forecasting and Control. Holden-Day, San Francisco.

Box, G.E.P, Tiao, G.C., 1975. Intervention analysis with applications to economic and environmental problems. Journal of the American Statistical association, 70 (349), 70–79. DOI: 10.2307/2285379.

Box, G.E.P., Jenkins, G.M., Reinsel, G.C., 1994. Time Series Analysis: Forecasting and Control. 3rd Edition. Prentice Hall, Englewood Cliff, New Jersey.

Ceylan, Z., 2020. Estimation of COVID-19 prevalence in Italy, Spain, and France. The Science of the Total Environment, 729, 138817. https://doi.org/10.1016/j.scitotenv.2020.138817.

Chakraborty, T., Ghosh, I., 2020. Real-time forecasts and risk assessment of novel coronavirus (COVID-19) cases: A data-driven analysis. Chaos, Solitons & Fractals, 135, 109850. https://doi.org/10.1016/j.chaos.2020.109850.

Cong, J., Ren, M., Xie, S., Wang, P., 2019. Predicting Seasonal Influenza Based on SARIMA Model, in Mainland China from 2005 to 2018. International Journal of Environmental Research and Public Health, 16 (23), 4760. doi: 10.3390/ijerph16234760.

Cryer, J.D., 1986. Time Series Analysis. Duxbury Press, Boston, Massachusetts.

Dickey, D.A., Fuller, W.A., 1979. Distribution of the Estimators for Autoregressive Time Series with a Unit Root. Journal of the American Statistical Association, 74 (366), 427–431. DOI: 10.2307/2286348.

Earnest, A., Chen, M.I., Ng, D., Sin, L.Y., 2005. Using autoregressive integrated moving average (ARIMA) models to predict and monitor the number of beds occupied during a SARS outbreak in a tertiary hospital in Singapore. BMC Health Services Research, 5 (1), 36. doi: 10.1186/1472-6963-5-36.

Fanelli, D., Piazza, F., 2020. Analysis and forecast of COVID-19 spreading in China, Italy and France. Chaos, Solitons & Fractals, 134, 109761. https://doi.org/10.1016/j.chaos.2020.109761.

Gaudart, J., Touré, O., Dessay, N., lassane Dicko, A., Ranque, S., Forest, L., Demongeot J., Doumbo, O.K., 2009. Modelling malaria incidence with environmental dependency in a locality of Sudanese savannah area, Mali. Malaria journal, 8 (61). https://doi.org/10.1186/1475-2875-8-61.

Giordano, G., Blanchini, F., Bruno, R., Colaneri, P., Di Filippo, A., Di Matteo, A., Colaneri, M., 2020. Modelling the COVID-19 epidemic and implementation of population-wide interventions in Italy. Nature Medicine. https://doi.org/10.1038/s41591-020-0883-7.

Goodwin, P., Lawton, R., 1999. On the asymmetry of the symmetric MAPE. International Journal of Forecasting, 15 (4), 405-408. https://doi.org/10.1016/S0169-2070(99)00007-2.

He, Z., Tao, H., 2018. Epidemiology and ARIMA model of positive-rate of influenza viruses among children in Wuhan, China: A nine-year retrospective study. International Journal of Infectious Diseases, 74, 61–70. https://doi.org/10.1016/j.ijid.2018.07.003.

Hurvich, C.M., Tsai, C.L., 1989. Regression and time series model selection in small samples. Biometrika, 76 (2), 297–307. https://doi.org/10.1093/biomet/76.2.297.

Hyndman, R.J., Athanasopoulos, G., 2018. Forecasting: Principles and Practice. Otext.

Hyndman, R.J., Khandakar, Y., 2008. Automatic time series forecasting: The forecast package for R. Journal of Statistical Software, 27 (1), 1–22. https://doi.org/10.18637/jss.v027.i03.

Hyndman, R.J., Koehler, A.B., 2006. Another look at measures of forecast accuracy. International journal of forecasting, 22 (4), 679–688. https://doi.org/10.1016/j.ijforecast.2006.03.001.

Kane, M.J., Price, N., Scotch, M., Rabinowitz, P., 2014. Comparison of ARIMA and Random Forest time series models for prediction of avian influenza H5N1 outbreaks. BMC bioinformatics, 15 (1), 276. https://doi.org/10.1186/1471-2105-15-276.

Kim, S., Kim, H., 2016. A new metric of absolute percentage error for intermittent demand forecasts. International Journal of Forecasting, 32 (3), 669–679. https://doi.org/10.1016/j.ijforecast.2015.12.003.

Kwiatkowski,, D, Phillips, P.C.B., Schmidt. P., Shin, Y., 1992. Testing the Null Hypothesis of Stationarity against the Alternative of a Unit Root. Journal of Econometrics, 54 (1–3), 159–178. https://doi.org/10.1016/0304-4076(92)90104-Y.

Lewis, C.D., 1982. Industrial and business forecasting methods. Butterworths, London.

Li, Q., Guo, N.N., Han, Z.Y., Zhang, Y.B., Qi, S.X., Xu, Y.G., Wei, Y.M., Han, X., Liu, Y.Y., 2012. Application of an autoregressive integrated moving average model for predicting the incidence of hemorrhagic fever with renal syndrome. The American journal of tropical medicine and hygiene, 87 (2), 364–370. DOI: https://doi.org/10.4269/ajtmh.2012.11-0472.

Liu, Q., Liu, X., Jiang, B., Yang, W., 2011. Forecasting incidence of hemorrhagic fever with renal syndrome in China using ARIMA model. BMC infectious diseases, 11 (1): 218. https://doi.org/10.1186/1471-2334-11-218.

McCleary, R., Hay, R.A., Meidinger, E.E., McDowall, D., 1980. Applied time series analysis for the social sciences. Sage Publications, Beverly Hills, CA.

Moreno, J.J.M., Pol, A.P., Abad, A.S., Blasco, B.C., 2013. Using the R-MAPE index as a resistant measure of forecast accuracy. Psicothema, 25 (4), 500–506. doi: 10.7334/psicothema2013.23.

Nesteruk, I., 2020. Statistics-Based Predictions of Coronavirus Epidemic Spreading in Mainland China. Innovative Biosystems and Bioengineering, 4 (1), 13–18. DOI: https://doi.org/10.20535/ibb.2020.4.1.195074.

Pack, D.J., 1990. In defense of ARIMA modeling. International Journal of Forecasting, 6 (2), 211–218. https://doi.org/10.1016/0169-2070(90)90006-W.

Phillips, P.C.B., Perron, P., 1988. Testing for a unit root in time series regression. Biometrika, 72 (2), 335–346. DOI: 10.2307/2336182.

Polwiang, S., 2020. The time series seasonal patterns of dengue fever and associated weather variables in Bangkok (2003-2017). BMC Infectious Diseases, 20 (1), 1–10. https://doi.org/10.1186/s12879-020-4902-6.

Ren, H., Li, J., Yuan, Z.A., Hu, J.Y., Yu, Y., Lu, Y.H., 2013. The development of a combined mathematical model to forecast the incidence of hepatitis E in Shanghai, China. BMC infectious diseases, 13 (1), 421. https://doi.org/10.1186/1471-2334-13-421.

Ren, L., Glasure, Y., 2009. Applicability of the revised mean absolute percentage errors (MAPE) approach to some popular normal and nonnormal independent time series. International Advances in Economic Research, 15 (4), 409-420. DOI: 10.1007/s11294-009-9233-8.

Ribeiro, M.H.D.M., da Silva, R.G., Mariani, V.C., dos Santos Coelho, L., 2020. Short-term forecasting COVID-19 cumulative confirmed cases: Perspectives for Brazil. Chaos, Solitons & Fractals, 135, 109853. https://doi.org/10.1016/j.chaos.2020.109853.

Rios, M., Garcia J.M., Sanchez J.A., Perez, D., 2000. A statistical analysis of the seasonality in pulmonary tuberculosis. European Journal of Epidemiology, 16 (5), 483–488. https://doi.org/10.1023/A:1007653329972.

Singh, R.K.., Rani, M., Bhagavathula, A.S., Sah, R., Rodriguez-Morales, A.J., Kalita, H., Nanda, C., Patairiya, S., Sharma, Y.D., Rabaan A.A., Rahmani, J., Kumar, P., 2020. Prediction of the COVID-19 Pandemic for the Top 15 Affected Countries: Advanced Autoregressive Integrated Moving Average (ARIMA) Model, JMIR Public Health Surveill, 6 (2), e19115. DOI: 10.2196/19115.

Sugiura, N., 1978. Further analysts of the data by akaike's information criterion and the finite corrections: Further analysts of the data by akaike's. Communications in Statistics-Theory and Methods, 7 (1), 13–26. https://doi.org/10.1080/03610927808827599.

Tuite, A.R., Ng, V., Rees, E., Fisman, D., 2020. Estimation of COVID-19 outbreak size in Italy. The Lancet infectious diseases, 20 (5), 537. DOI: 10.1016/S1473-3099(20)30227-9.

Wang, L., Liang, C., Wu, W., Wu, S., Yang, J., Lu, X., Cai, Y., Jin, C., 2019. Epidemic Situation of Brucellosis in Jinzhou City of China and Prediction Using the ARIMA Model. Canadian Journal of Infectious Diseases and Medical Microbiology, 1429462. https://doi.org/10.1155/2019/1429462.

Wang, Y.W., Shen, Z.Z., Jiang, Y., 2018. Comparison of ARIMA and GM (1, 1) models for prediction of hepatitis B in China. PloS One, 13 (9): e0201987. doi:10.1371/journal.pone.0201987,

Wei, W., Jiang, J., Liang, H., Gao, L., Liang, B., Huang, J., 2016. Application of a Combined Model with Autoregressive Integrated Moving Average (ARIMA) and Generalized Regression Neural Network (GRNN) in Forecasting Hepatitis Incidence in Heng County, China. PLoS ONE, 11 (6), e0156768. doi: 10.1371/journal.pone.0156768.

Wu, J.T., Leung, K., Leung, G.M., 2020. Nowcasting and forecasting the potential domestic and international spread of the 2019-nCoV outbreak originating in Wuhan, China: a modelling study. The Lancet, 395 (10225), 689–697. https://doi.org/10.1016/S0140-6736(20)30260-9.

Xiong, H., Yan, H., 2020. Simulating the infected population and spread trend of 2019-nCov under different policy by EIR model, medRxiv. doi: https://doi.org/10.1101/2020.02.10.20021519.

Xu, Q., Li, R., Liu, Y., Luo, C., Xu, A., Xue, F., Xu, Q., Li, X., 2017. Forecasting the incidence of mumps in Zibo City based on a SARIMA model. International journal of environmental research and public health, 14 (8), 925. doi: 10.3390/ijerph14080925.

Zeng, Q., Li, D., Huang, G., Xia, J., Wang, X., Zhang, Y., Tang, W., Zhou, H., 2016. Time series analysis of temporal trends in the pertussis incidence in Mainland China from 2005 to 2016. Scientific reports, 6 (1), 1–8. https://doi.org/10.1038/srep32367.

Zhang X., Zhang T., Young A. A., Li X (2014), Applications and Comparisons of Four Time Series Models in Epidemiological Surveillance Data, PLoS One, 9 (2), e91629. https://doi.org/10.1371/journal.pone.0091629.

Zhang, X., Ma, R., Wang, L., 2020. Predicting turning point, duration and attack rate of COVID-19 outbreaks in major Western countries. Chaos, Solitons & Fractals, 135, 109829. https://doi.org/10.1016/j.chaos.2020.109829.

Zheng, Y.L., Zhang, L.P., Zhang, X.L., Wang, K., Zheng, Y J., 2015. Forecast model analysis for the morbidity of tuberculosis in Xinjiang, China. PloS one, 10 (3), e00116832. https://doi.org/10.1371/journal.pone.0116832.

APPENDIX A

Table A1. Mathematical equations for forecasting accuracy measures.

| Forecasting accuracy measures | Formula |
|---|---|
| MAE | $\frac{1}{n}\sum_{i=1}^{n}|y_i - \hat{y}_i|$ |
| MAPE | $\frac{1}{n}\sum_{i=1}^{n}\frac{|y_i-\hat{y}_i|}{y_i}*100\%$ |
| MASE | $\frac{1}{n}\sum_{i=1}^{n}\left(\frac{|y_i - \hat{y}_i|}{\frac{1}{n-1}\sum_{i=2}^{n}|y_i - \hat{y}_i - 1|}\right)$ |
| RMSE | $\sqrt{\frac{1}{n}\sum_{i=1}^{n}(y_i - \hat{y}_i)^2}$ |

Notes: $y_i$, actual values; $\hat{y}_i$, predicted values.

APPENDIX B

Table B1. Estimated parameters for the ARIMA models.

| Parameters | Italy | Russia | USA |
|---|---|---|---|
| AR (1) | 0.9029*** | -0.2732** | -0.8784*** |
|  | [0.1518] | [0.1383] | [0.1507] |
| AR (2) | -1.2004*** |  | -0.4945*** |
|  | [0.1836] |  | [0.1722] |
| AR (3) | 0.4698*** |  | -0.8459*** |
|  | [0.1781] |  | [0.1094] |
| AR (4) | -0.5392*** |  | -0.8046*** |
|  | [0.1316] |  | [0.1024] |
| AR (5) |  |  | -0.7234*** |
|  |  |  | [0.1494] |
| AR (6) |  |  | -0.4981*** |
|  |  |  | [0.1248] |
| MA (1) | -2.0183*** | -0.8517*** | -0.1967 |
|  | [0.1428] | [0.0924] | [0.1577] |
| MA (2) | 2.3886*** |  | -0.4594*** |
|  | [0.279] |  | [0.1365] |
| MA (3) | -1.751*** |  | 0.4931*** |
|  | [0.3091] |  | [0.1459] |
| MA (4) | 0.725*** |  |  |
|  | [0.1595] |  |  |

***p-value < 0.01; **p-value < 0.05. Standard errors in brackets.

Table B2. The results of the Ljung Box's test for autocorrelation (Figures 4, 5, and 6).

| Country | Ljung Box test for autocorrelation | | |
|---|---|---|---|
| | Statistics | p-value | Decision |
| Italy | | | |
| Lags (T/4= 13.25) | 3.3837 | 0.6411 | No autocorr. |
| Lags (12) | 3.3837 | 0.4958 | No autocorr. |
| Lags ($\sqrt{T} + 10 = 17.28$) | 4.0238 | 0.9098 | No autocorr. |
| Lags (20) | 5.9601 | 0.9181 | No autocorr. |
| Lags (10) | 3.0176 | 0.2212 | No autocorr. |
| | | | |
| Russia | | | |
| Lags (T/4=15.5) | 12.815 | 0.5411 | No autocorr. |
| Lags (12) | 10.11 | 0.4309 | No autocorr. |
| Lags ($\sqrt{T} + 10 = 17.87$) | 12.8178 | 0.686 | No autocorr. |
| Lags (20) | 13.726 | 0.7468 | No autocorr. |
| Lags (10) | 9.0948 | 0.3344 | No autocorr. |
| | | | |
| USA | | | |
| Lags (T/4 = 17.25) | 6.2393 | 0.6205 | No autocorr. |
| Lags (12) | 3.9367 | 0.2684 | No autocorr. |
| Lags ($\sqrt{T} + 10 = 18.31$) | 6.4613 | 0.693 | No autocorr. |
| Lags (20) | 9.3568 | 0.589 | No autocorr. |
| Lags (10) | 2.1681 | 0.1409 | No autocorr. |

Notes: lag selection (H) is based on Box and Jenkins (1970) (H=T/4), Bowerman and O'Connell (1979) (H=12), Cryer (1986) (H=$\sqrt{T}$ + 10), Shumway and Stoffer (2011) (H=20), and Hyndman and Athanasopoulos (2018) (H=10).

Table B3. The results of Engle's LM test for the ARCH effect (Figures 4, 5, and 6).

| Countries | Engle's LM test for the ARCH effect | | |
|---|---|---|---|
| | Statistics | p-value | Decision |
| Italy | | | |
| Lag (1) | 2.0166 | 0.1556 | No arch effect |
| Lag (12) | 10.9978 | 0.5291 | No arch effect |
| Lag (24) | 25.2823 | 0.3906 | No arch effect |
| Russia | | | |
| Lag (1) | 3.5742 | 0.0587 | No arch effect |
| Lag (12) | 11.2262 | 0.5096 | No arch effect |
| Lag (24) | 14.7925 | 0.871 | No arch effect |
| USA | | | |
| Lag (1) | 0.0478 | 0.827 | No arch effect |
| Lag (12) | 2.7772 | 0.9969 | No arch effect |
| Lag (24) | 14.9558 | 0.922 | No arch effect |

APPENDIX C

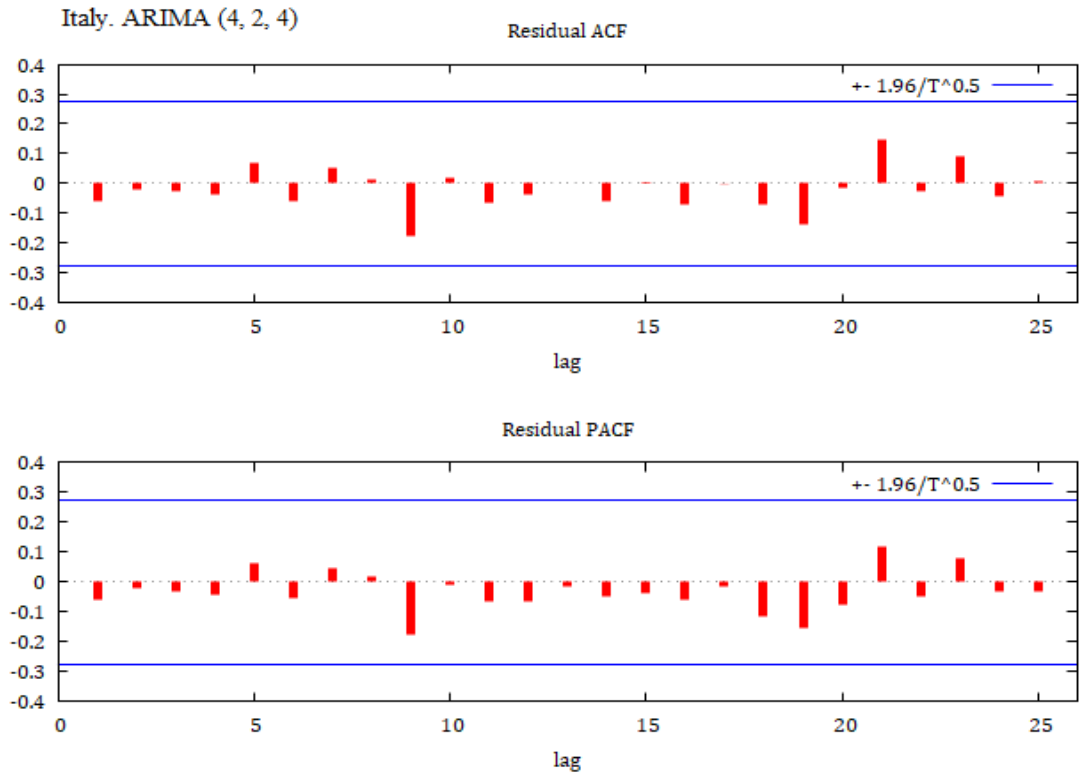

Figure C1. ACF and PACF plots of residuals for Italy.

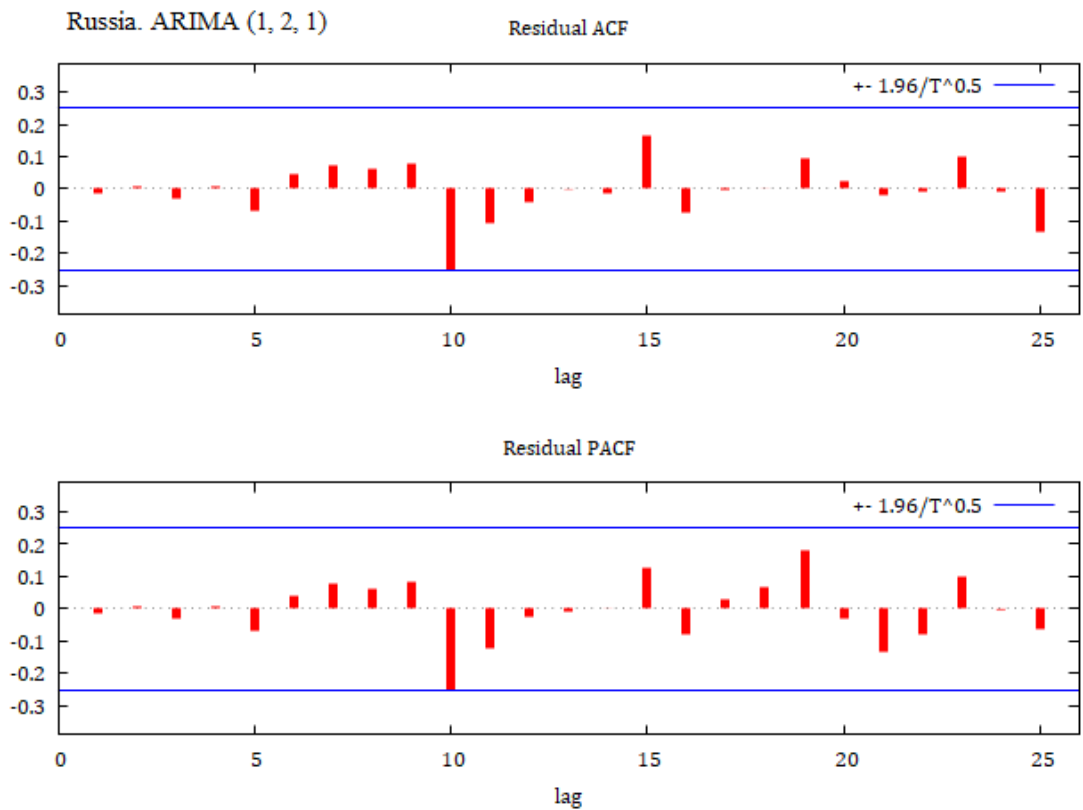

Figure C2. ACF and PACF plots of residuals for Russia.

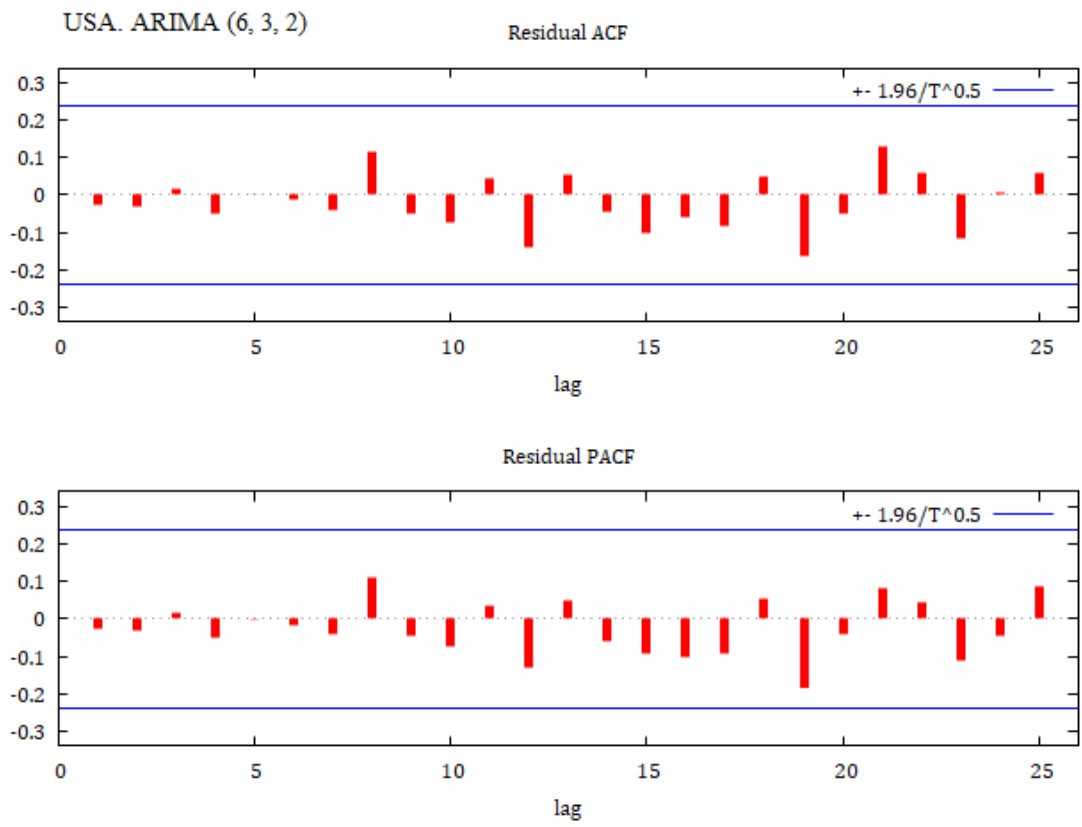

Figure C3. ACF and PACF plots of residuals for the USA.